# Trend Investigation of Biopotential Recording Front-End Channels for Invasive and Non-Invasive Applications

Taeju Lee[1] and Minkyu Je[2]

[1]Department of Electrical Engineering, Columbia University, New York, NY 10027, USA
E-mail: taeju.lee@columbia.edu

[2]School of Electrical Engineering, Korea Advanced Institute of Science and Technology, Daejeon 34141, Korea
E-mail: mkje@kaist.ac.kr

*Abstract*—Over the past decades, neurotechnology has made significant progress along with remarkable advances in microfabrication technology. Brain activity has been observed in a variety of modalities such as electrical, optical, and chemical recordings. Signals recorded by each modality exhibit different properties in magnitude and frequency, thereby requiring an appropriate recording front-end topology for each modality. For obtaining meaningful information in a complex neural network, brain activity can be observed by using multiple modalities rather than using a single modality. However, among these modalities, electrical recording has been more widely used for tracking neuron activity than other modalities. The electrical recording is carried out using the probe and recording front-end channel. The probe detects the weak voltage signals from neurons and the recording front-end channel processes the voltage signal detected via the probe. Thanks to advances in microfabrication technology, the probes have been developed in diverse physical forms while incorporating a large number of electrode sites, and front-end channels have also been developed to have various physical forms and functions. This paper presents the trend of biopotential recording front-end channels developed from the 1970s to the 2020s while describing a basic background on the front-end channel design. Only the front-end channels that conduct electrical recording invasively and non-invasively are addressed. The front-end channels are investigated in terms of technology node, number of channels, supply voltage, noise efficiency factor, and power efficiency factor. Also, multi-faceted comparisons are made to figure out the correlation between these five categories. In each category, the design trend is presented over time, and related circuit techniques are discussed. While addressing the characteristics of circuit techniques used to improve the channel performance, what needs to be improved is also suggested. The following is the brief table of contents for this paper.

TABLE OF CONTENTS

## I. INTRODUCTION

The interfacing with the brain has made tremendous achievements in electrophysiology and neurophysiology. Understanding of brain activity has been conducted through various probe types according to the signal observing site. Also, the interfacing with neuronal cells by culturing them on the microelectrode array has enabled studies for cell assessment, imaging, drug development, and long-term neural activity analysis. Although this is not the only way, generally, the monitoring of neuronal signals can be carried out through the amplification stage and analog-to-digital converter (ADC). The digital output signals from the ADC are relayed to the digital interface for data transmission, signal processing, and data interpretation.

Thanks to the advancement of semiconductor technology, biopotential recording front-end integrated circuits (ICs) have been developed in various physical forms with advanced performance. Fig. 1 summarizes the widely employed biopotential recording front-end ICs that are classified into two domains according to their physical form. One is a general purpose (GP) system which includes the front-end IC connected with the passive probe through the cable, the active probe incorporating the front-end circuits and electrode sites on the same substrate, and the active probe with the pixel amplifiers placed under each electrode site. The other form is a microelectrode array (MEA) system which is designed by incorporating the electrode array and front-end circuits on the same substrate for conducting diverse *in vitro* studies.

Throughout this paper, the GP and MEA systems are investigated in terms of five categories such as technology node, number of front-end channels, supply voltage, noise efficiency factor, and power efficiency factor based on the previous research works [1]–[358]. The prior works [1]–[358], selected based on the silicon fabrication and experiment results, include invasive and non-invasive biopotential recording front-end channels developed from the 1970s to the 2020s. The biopotential signals can be observed through modalities of electrical, optical, and chemical recordings. In this paper, only the front-end channels that conduct electrical recording are addressed. In the modality of electrical recording, various voltage signals are observed depending on the signal monitoring site through the probe. The observable voltage signals include action potentials (APs), local field potentials (LFPs), electrocorticography (ECoG) signals, and ExG signals. The ExG refers to electroencephalogram (EEG), electrocardiogram (ECG), electromyogram (EMG), and electrooculogram (EOG). Throughout this paper, the data points in each figure are noted as orange and yellow circles to express the GP and MEA systems, respectively. Along with the trend investigation of each category from the past to the present, multi-faceted comparisons are conducted to figure out the correlation between the five categories. Also, the trend investigation of each category discusses the key circuit techniques used in the front-end channel design and suggests what needs to be improved.

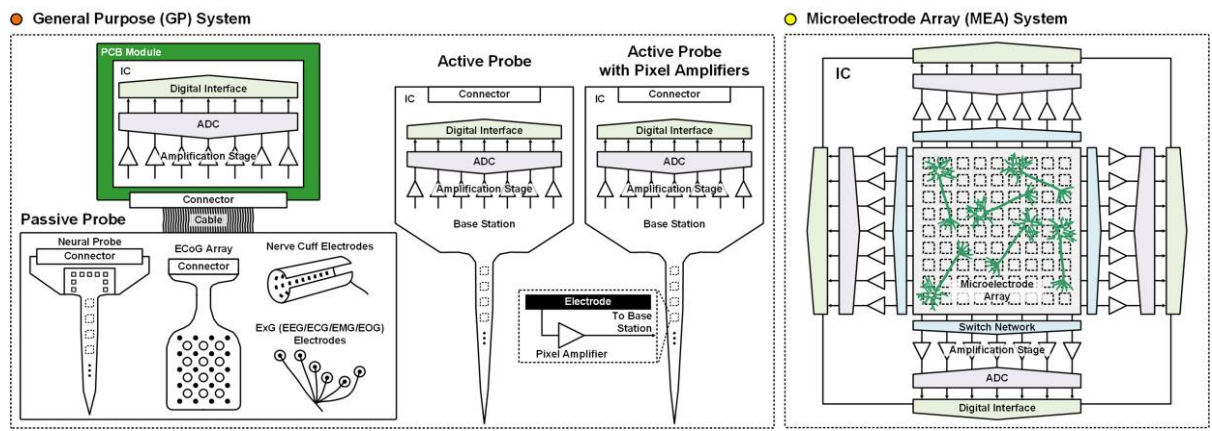

Fig. 1. Biopotential recording front-end ICs according to the physical form.

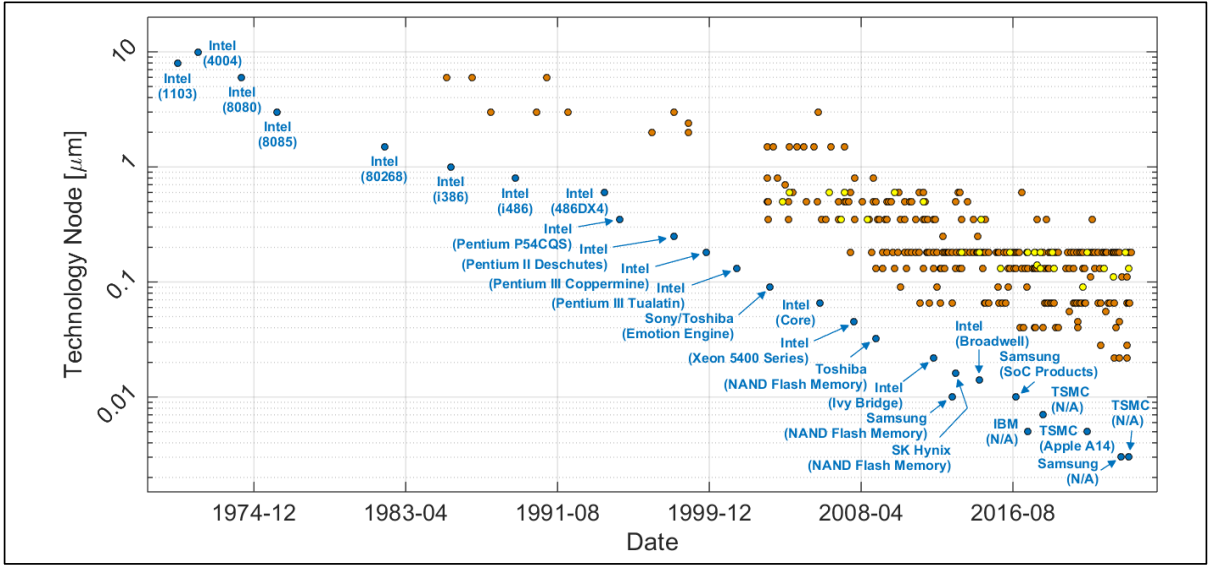

Fig. 2. Technology node trends of biopotential recording front-end channels and commercial products developed from the 1970s to the 2020s (orange circles: GP system; yellow circles: MEA system; sky circles: commercial products).

## II. TECHNOLOGY NODE TREND OF BIOPOTENTIAL RECORDING FRONT-END CHANNELS

The advent of semiconductor technology has triggered a tremendous impact on our life, thereby developing computers, cellular phones, wireless devices, mobile devices, etc. As semiconductor technology advances, the operating speed of devices becomes faster and their physical form is minimized while integrating a huge number of transistors. As the technology node shrinks, the battery lifetime of the devices is prolonged by consuming low power. Therefore, advances in technology nodes have enabled the development of high-performance electronics in various fields such as consumer electronics, military, space, biomedical industries, etc. In particular, the advancement of semiconductor technology has led to significant progress in biomedical applications that need to satisfy conditions of low-power operation and miniaturization. Among various biomedical applications, the biopotential recording front-end channels have evolved along with the advancement of the technology node, thereby making remarkable achievements in electrophysiology and neurophysiology.

Fig. 2 presents the technology node trends of biopotential recording front-end channels along with the commercial products. The technology node trend of commercial products is investigated from the 1970s to the 2020s and indicated as sky circles. The technology node trend of the GP systems is expressed using the orange circles while the MEA systems are expressed as yellow circles. Note that the trends of GP and MEA systems are investigated since the 1970s but some data points are missing from the 1970s to the 1980s in Fig. 2 because the used technology node is not mentioned in the literature.

As shown in the trend of commercial products (Fig. 2), the technology node has shrunk by several leading manufacturers. As the technology node shrinks, electronic devices have become faster and more functional by increasing the integration density of transistors. Moreover, the power consumption of devices is reduced, thereby prolonging the battery lifetime and reducing heat generation. Therefore, the manufacturers have been trying to push their technology nodes to limits increasingly. Although commercial products and biopotential recording front-end channels show a time gap between the two trends as shown in Fig. 2, the biopotential recording front-end channels show a similar trend to that of commercial products. Compared to the period from the 1970s to the 1990s when the front-end channels are rarely developed, numerous front-end ICs for biopotential recording have been developed since the 2000s using various technology nodes (Fig. 2). Especially, the 180 nm node has been steadily used for channel development for a long period compared to other nodes.

*A. Design Characteristics for Biopotential Recording Systems*

Compared to digital-intensively designed commercial products such as the central processing unit (CPU) and graphics processing unit (GPU), many biopotential recording systems have been designed analog-intensively. The main role of the recording system is to sense external biopotential signals and digitize the sensed signals. In addition, in some cases, the recording system needs to be designed with a large number of front-end channels for obtaining richer biological information. However, a certain amount of area occupied by analog circuits is required for designing a high-performance front-end channel. Therefore, the channels can occupy a large portion of the design area in a biopotential recording system. Given that the electrical performance of the digital-intensive system improves as the technology node shrinks, the biopotential recording system may not benefit from technology scaling as much as the digital-intensive circuitry such as the CPU and GPU.

To ensure reliable operation of the recording system in practical use such as *in vivo* and *in vitro* studies, each front-end channel needs to be designed considering the parameters such as the input-referred noise, power consumption, nonlinearity, etc. Among the various parameters, the input-referred noise and power consumption are primary concerns in a design. For detecting weak biopotential signals, the input-referred noise must be sufficiently low. The power consumption must be minimized to prevent tissue damage due to heat generation. However, it cannot be said that technology scaling always brings out the best performance in the input-referred noise and power consumption. In addition to the low noise and low power performance, the front-end channels must be designed to have good linearity to avoid signal distortion.

When designing the biopotential recording front-end channel, the thermal and flicker noise generated from the channel must be achieved as low as possible for detecting weak voltage signals. For achieving a good thermal noise efficiency, the relative transconductance $g_m/I_D$ of the input transistor used in the front-end channel needs to increase [20], where $g_m$ and $I_D$ are the transconductance and bias current of the input transistor, respectively. A high $g_m/I_D$ can be achieved by maximizing the $W/L$ ratio of the input transistor [20], where $W$ and $L$ are the width and length of the input transistor. The flicker noise can be minimized by achieving a large gate area $WL$ of the input transistor [359]. As a result, a high $W/L$ ratio and a large area of $WL$ are required for minimizing the input-referred thermal and flicker noise. In addition to the transistor size, a sufficient bias current needs to be supplied to the input transistors for increasing $g_m$. In other words, rather than using the small design area and low bias current, a certain amount of the area and bias current needs to be used to achieve the best performance while balancing between the noise and power consumption.

Considering these design conditions and fabrication costs, it may be a reasonable design approach to use the mid-scale technology nodes instead of the state-of-the-art technology nodes if the analog circuitry occupies a large portion of the area than the digital circuitry in a total system. As shown in Fig. 2, the 180 nm node has been widely used to implement the front-end channels for a long period among the various technology nodes. However, as an alternative design approach to overcome the area inefficiency of the analog-intensive front-end channel and improve the performance, the channel can be designed digital-intensively by reducing the dependence on analog circuits. For this reason, the need for advanced technology nodes such as 65 nm, 40 nm, etc. is increasing (Fig. 2).

*B. Widely Employed Recording Front-End Architectures*

Since the 1970s, biopotential recording front-end channels have been developed using various technologies. In the early stage of technology development, the front-end channels are implemented using the junction-gate field-effect transistor (JFET), liquid-oxide-semiconductor field-effect transistor (LOSFET), p-channel metal-oxide-semiconductor (MOS), and n-channel MOS technologies [1]–[6], [9]. These front-end channels developed using older technologies are designed as the source follower and open-loop amplifier.

As semiconductor technology matures, complementary metal-oxide semiconductor (CMOS) technology has been widely used in the design of biopotential recording front-end channels since the 1990s. To set the conversion gain of the front-end channels to be robust to process variations, the channels can be designed using various feedback networks such as current feedback, resistive feedback, and capacitive feedback. Among these feedback networks, the capacitive feedback structure has been widely employed for the design of low-noise low-power amplifiers [20]. Compared to the current and resistive feedback-based amplifiers [25], [40], the capacitive feedback-based amplifier can filter out the rail-to-rail DC voltage by connecting a capacitor to its input. Therefore, DC electrode offsets generated as a result of the electrochemical response between the electrode surface and the tissue (or skin) can be effectively removed without external components. Numerous capacitive feedback-based amplifiers have been implemented using various CMOS technology nodes, and their structures have evolved to improve the performance, e.g., the input-referred noise, input impedance, and dynamic range.

Especially, when monitoring ECoG and EEG signals, it is necessary to reduce the flicker noise because the signal energy is distributed in the low-frequency band. Several methods such as autozeroing, chopping, and switched biasing can be used to reduce the flicker noise [360]–[362], but among them, the chopping technique has been widely employed in the low-noise amplifier design. The principle of the chopping technique is to up-modulate the input signal with a carrier frequency and amplify the modulated signal by avoiding the flicker noise band. Then, the signal is reconstructed by demodulating the amplified signal. When applying the chopping technique to the capacitive feedback-based amplifier, the flicker noise can be effectively filtered out. But the DC offset at the amplifier input is also up-modulated by the chopper along with the biopotential signal. Accordingly, the amplifier output can be saturated when the offset voltage becomes large. To suppress the DC offset-modulated signal, the DC servo loop (DSL) must be employed in the amplifier [90]. In addition to the DSL, for high-precision operation, the ripple reduction loop (RRL) is required to suppress the chopping ripple due to the offset of the operational transconductance amplifier (OTA) used in the capacitive feedback-based amplifier [90]. However, when employing both the DSL and RRL in a single channel, the design can be complicated while entailing side effects caused by the RRL. Therefore, as an alternative way to alleviate the chopping ripple without additional power consumption, a DC-blocking capacitor can be placed inside the OTA [203].

The chopper is a good solution for suppressing the flicker noise, but the input impedance of the front-end channel needs to be considered when using the chopping technique. The input impedance of the front-end channel is proportionally degraded to the chopping frequency $f_{chop}$ and input capacitor $C_{in}$. Considering the high source impedance of the probe, the low input impedance causes signal attenuation between the probe and the channel. Also, the common-mode rejection ratio (CMRR) of the channel is degraded due to the low input impedance. To mitigate these issues, various circuit techniques are developed. Among them, the positive feedback loop and auxiliary path are widely employed for boosting the input impedance [90], [203]. A T-capacitor feedback network can be employed in the capacitive feedback-based amplifier to increase the input impedance using a smaller $C_{in}$ [139]. However, these boosting techniques can entail a degradation in the performance, e.g., the input-referred noise and stability, so the techniques can be applied in a limited range. Also, instead of using an AC-coupled structure such as the capacitive feedback-based amplifier, the front-end channel can be designed as a DC-coupled structure where the input signal is directly applied to the gate terminal of the metal-oxide-semiconductor field-effect transistor (MOSFET), thereby increasing the input impedance by removing $C_{in}$ [104]. However, the input impedance of a DC-coupled structure is still affected by the parasitic components of the input MOSFET.

In addition to the circuit techniques described above, various circuit techniques have been developed to improve the performance of the front-end channels. But each channel can occupy a fairly large design area because multiple OTAs, large capacitors, and high resistance may be required for achieving the best figure of merit. Nevertheless, the analog-intensive front-end channel is a good solution for ensuring reliable operation using mid-scale technology nodes. If a small number of recording channels are required, the analog-intensive front-end channel is still an attractive option considering the performance and fabrication costs. Therefore, the mid-scale technology nodes have been widely used so far for various biopotential recording systems as shown in Fig. 2.

*C. Digital-Intensive Recording Front-End Architectures using Advanced Technology Nodes*

When designing the front-end channel based on the aforementioned techniques, the area of each channel can be quite increased, thereby limiting the implementation of the multi-channel recording system. Conventionally, numerous multi-channel recording systems are implemented using the low-noise amplifier (LNA), variable gain amplifier (VGA), and ADC. Note that the VGA can be also expressed as the programmable gain amplifier (PGA) in the literature and both are used interchangeably. The detected biopotential signals through the LNA and VGA are digitized by the ADC. For an area-efficient design, multiple LNAs and VGAs can share one ADC through the multiplexer, and the ADC operates at a high sampling rate. When designing the LNA and VGA as the capacitive feedback-based amplifier, the overall area of the recording system can be significantly increased as the number of recording channels increases. Therefore, the area-efficient design of each recording channel becomes important in the design of the multi-channel recording system.

For overcoming the area inefficiency of the analog-intensive front-end channel consisting of the LNA, VGA, and ADC, the direct conversion channel is widely used. The direct conversion channel can be implemented using various circuit topologies. But the most commonly used circuit structure is composed of the integrator, quantizer, and feedback digital-to-analog converter (DAC). The integrator, quantizer, and feedback DAC form a closed-loop system, then the biopotential signal is directly digitized through the closed-loop feedback operation without passing through the LNA and VGA. This is similar to the operation of the delta-sigma (ΔΣ) modulator. There are several advantages of using the direct conversion channel compared to the analog-intensive front-end channel consisting of the LNA, VGA, and ADC as follows:

(1) The direct conversion channel does not require multiple amplification stages, e.g., the LNA and VGA, consisting of multiple OTAs and large capacitors. Therefore, the area increase can be avoided due to multiple amplification stages and an area-efficient channel design can be possible. It means that the direct conversion channel can be extended into a large number of recording channels.

(2) The direct conversion channel can be designed more digital-intensively than the LNA and VGA. Accordingly, the channel can benefit from technology scaling. It means that when the advanced technology node is used, the direct conversion channel can be area- and power-efficiently designed. Eventually, the channel is suitable for extension into an energy-efficient multi-channel recording system.

(3) In the recording channel consisting of the LNA, VGA, and ADC, the biopotential signal is first amplified with a high voltage gain by the LNA and VGA. Then, the amplified signal is digitized by the ADC. Therefore, the dynamic range of the whole recording chain can be degraded. However, the direct conversion channel can achieve a wide dynamic range by avoiding the high gain stages and having noise-shaping characteristics.

Thanks to these advantages, the various direct conversion channels have been developed using advanced technology nodes such as 65 nm, 55 nm, 40 nm, and 22 nm as shown in Fig. 2 [104], [163], [199], [211], [230], [235], [244], [263], [265], [294], [298], [299], [304], [311], and [321]. Although the portion of digital circuits increases compared to the LNA and VGA while reducing the portion of analog circuits, the input-referred noise performance of the direct conversion channel is still degraded by analog circuits such as the integrator. So the chopping technique can be applied and the size of the input devices should not be designed too small.

Since the early 1990s, various biopotential recording channels have been developed using standard CMOS technologies. Given the design conditions, e.g., the channel topology, voltage range, and operating frequency, appropriate semiconductor technology is employed for circuit design. When the recording channels are designed in conjunction with the stimulation circuits in the neuromodulation system, the channels need to be designed to have a wide dynamic range to tolerate stimulation artifacts while observing the biopotential signals with minimal distortion. The recording channels can be designed using high-voltage CMOS (HV-CMOS) and bipolar-CMOS-DMOS (BCD) technologies to tolerate the high-voltage signals induced by the stimulation circuits. Especially, when using HV-CMOS and BCD technologies, the recording channels can be protected from high-voltage signals by using high-voltage transistors. But when the recording channels are designed along with stimulation circuits using standard CMOS technology that does not support high-voltage transistors, the proper circuit techniques must be implemented using standard transistors to protect the channels from high-voltage signals. The protection of recording channels from high-voltage signals affects the channel lifetime and must be considered in the design.

### III. TREND OF THE NUMBER OF ACTIVE FRONT-END CHANNELS

For understanding the whole brain function, a large number of neurons must be detected and their signals need to be processed simultaneously. The number of simultaneously recorded neurons has gradually increased over the past decades [363], [364]. But the number of simultaneously recorded neurons is still insufficient to understand the whole brain of the mouse, monkey, and human [364]. For increasing the number of simultaneously observed neurons, a large number of front-end channels is essential along with a large number of electrode sites. Fig. 3 presents the number of active front-end channels fabricated using various technology nodes from the 1970s to the 2020s. Note that, in the case of the active probe with pixel amplifiers in Fig. 1, the number of channels is counted as the number of pixel amplifiers, e.g., source followers, common-source stages, and closed-loop amplifiers. Thanks to the advances in technology nodes and design techniques, the number of active front-end channels has gradually increased since the 2000s. As shown in Fig. 3, the number of channels developed until the end of the 1990s does not exceed twenty. But the number of channels developed from the early 2000s to the present has varied from one to tens of thousands. Note that the data points of Fig. 3 do not mean the number of simultaneously used channels but mean the total number of physically implemented front-end channels in the first sensing stage. The orange and yellow circles indicate GP and MEA systems, respectively.

*A. Design Strategies for High-Density Recording Systems*

To increase the density of recording channels in a given design area, the single front-end channel must be area-efficiently designed while minimizing the performance degradation in the input-referred noise and power consumption. Since the 2000s, for implementing the high-density multi-channel recording system, various design techniques have been developed while minimizing the area of the single front-end channel and maximizing the number of input signals that can be processed simultaneously. In most of the multi-channel recording systems, the differential structure has been widely used in the front-end channels for achieving high CMRR and high power supply rejection ratio (PSRR). So numerous multi-channel recording systems have been implemented with as many differential channels as the number of electrode sites. However, as the number of electrode sites increases, the overall area of the whole recording system can also increase significantly.

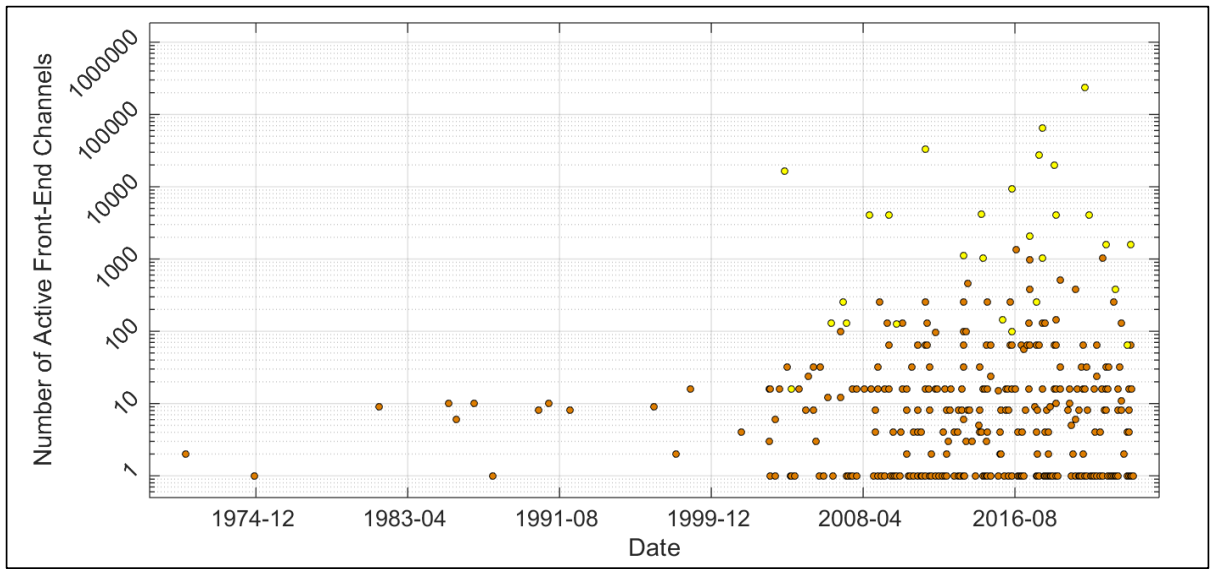

Fig. 3. Trend of the number of active front-end channels from the1970s to the 2020s.

To overcome the area inefficiency while maintaining the differential properties of channels, the multi-channel recording system can be designed as a pseudo-differential structure [146], [193], and [209]. Compared to the differential recording systems where the differential channels are placed in the first sensing stage, the pseudo-differential recording systems employ the single-ended channels in the first sensing stage. Among the multiple single-ended channels, an active channel is selected through the multiplexer for differential signal processing, and the remaining channels are time-multiplexed for scanning all electrode sites. The selected active single-ended channel is then further processed through the differential channel in the second sensing stage along with the reference single-ended channel. Thanks to the use of a pseudo-differential structure, the design area of the first sensing stage can be halved compared to a differential recording system. Also, the input impedance mismatch between the active and reference input ports can be mitigated compared to that of the differential recording systems where all reference input ports of differential channels are connected in the first sensing stage.

In addition to a pseudo-differential structure, for observing biopotential signals from the multiple electrode sites while occupying a small design area, the single front-end channel can operate in conjunction with the pre-summing circuit through the time-division multiple access (TDMA), frequency-division multiple access (FDMA), and code-division multiple access (CDMA) methods. The multiple access-based recording can be conducted by using the single front-end channel and pre-summing circuit that combines all input signals into the channel input. Therefore, these methods allow for the area-efficient system design compared to the differential recording systems where the overall area increases with the number of electrode sites. Also, the power consumption is reduced as much as the reduced channels, thereby enabling the energy-efficient design in multi-channel recording systems.

The TDMA-based front-end channel is implemented by connecting the multiple electrode sites to the input of the single front-end channel through the analog multiplexer that corresponds to the pre-summing circuit in the time domain [236], [305], [322], and [325]. For the single front-end channel to access multiple electrode sites with minimal loss of input signals, a high multiplexing frequency is required. Accordingly, the front-end channel must be designed to have a wide bandwidth. However, the DC offsets of electrode sites are up-modulated by the multiplexing frequency, and the modulated DC offsets are recorded along with the biopotential signals. It means the front-end channel needs to be designed to have a wide dynamic range. Also, the DC offset suppression technique can be required for high-precision signal processing in the TDMA-based front-end channel. When the channel accesses all electrode sites through an analog multiplexer, the charge injection into the channel input is induced depending on the multiplexing frequency and the number of electrode sites, lowering the input impedance of the front-end channel. Therefore, the impedance boosting technique is required for high-precision signal observation using the TDMA method.

The FDMA-based front-end channel is implemented by summing the frequency-modulated input signals into the input of the single front-end channel [167], [254]. The frequency modulation for each input signal is conducted by the chopper. As the number of input signals increases, each input signal is up-modulated using a different chopping frequency. Then, the signals are reconstructed at the output of the front-end channel through the demodulation using the choppers. For processing multiple frequency-modulated input signals with minimal signal loss, the single front-end channel must have a wide bandwidth. Also, for mitigating the inherent issues caused by the chopper, compensation techniques, e.g., the DSL, RRL, and input impedance boosting loop, may be required. Especially, the input signals can experience different input impedances since each input signal is up-modulated using a different chopping frequency. It means the signal attenuation of each input signal is different according to the input chopper and the attenuation imbalance between all input signals becomes severe as the number of input choppers increases. Accordingly, the input impedances of all input ports need to be properly boosted and equalized to avoid attenuation imbalance due to the increase in the number of input choppers.

The CDMA-based front-end channel can be designed by summing the orthogonally-modulated input signals into the input of the single front-end channel [276]. Each input signal is modulated using the orthogonal code by the chopper and merged into the input of the single front-end channel with the other orthogonally-modulated input signals. Then, the signal is reconstructed using the orthogonal code by the output chopper. Similar to the FDMA-based front-end channel, the CDMA-based front-end channel also can be implemented using the chopper at the input port, thereby inducing the issues of the DC offset modulation and input impedance degradation. Accordingly, appropriate compensation techniques, e.g., the DSL, RRL, and input impedance boosting loop, may be required

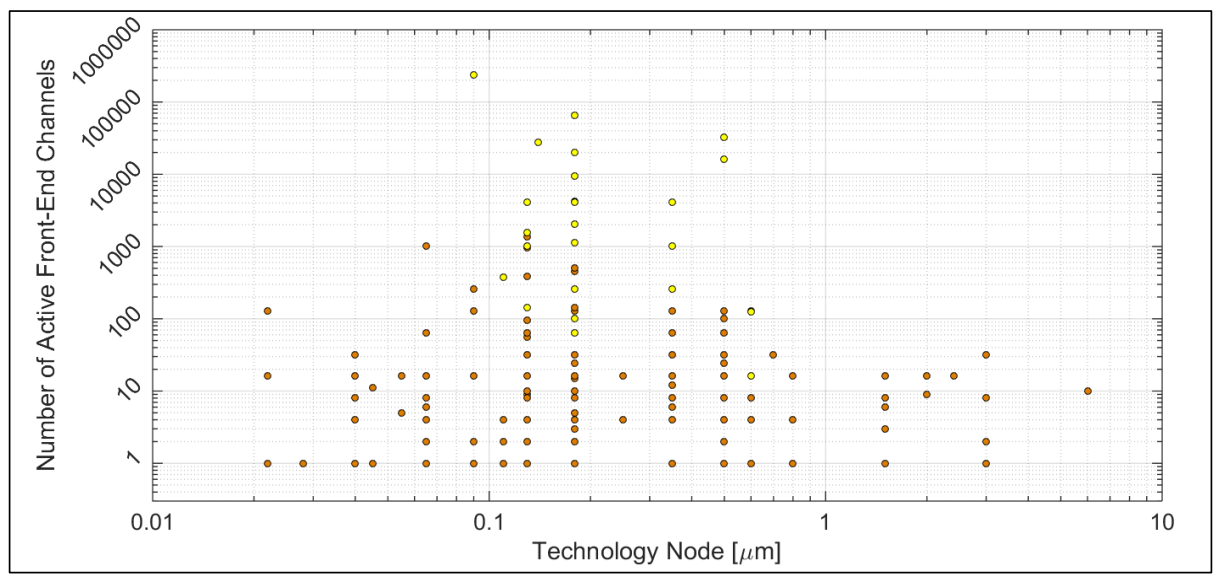

Fig. 4. The number of active front-end channels according to the used technology node.

for observing all input signals with minimal signal distortion.

As discussed above, TDMA-, FDMA-, and CDMA-based front-end channels can process multiple input signals by sharing one recording channel, enabling area-efficient design. However, these approaches have inherent issues such as the input DC offset modulation, power increase due to wide bandwidth, input impedance degradation due to the chopper, and design complexity. In particular, the performance degradation and design complexity can become severe when the number of input ports increases. Considering the general way of electrode connection in the two-port input such as the chopper, the reference ports of all choppers are connected all together for the reference signal, and the remaining active ports of all choppers are connected to electrode sites. In this case, the impedance mismatch between the reference and active ports can become greater as the number of choppers increases, especially in FDMA- and CDMA-based front-end channels, requiring the input impedance compensation technique optimized for each input port. Therefore, in the channel topologies, e.g., TDMA, FDMA, and CDMA methods, appropriate compensation techniques need to be chosen by considering trade-offs between the number of input ports, the whole design area, the performance, and design complexity.

*B. Number of Active Front-End Channels versus Technology Node*

Based on the biopotential recording front-end ICs developed from the 1970s to the 2020s, Fig. 4 presents the number of active front-end channels according to the used technology node. Among the various technology nodes, many multi-channel recording systems are implemented using the mid-scale technology nodes such as 130 nm and 180 nm. In addition, the nodes of 130 nm and 180 nm are widely used for implementing the biopotential recording front-end channels over the entire period as shown in Fig. 2. The increasing importance of the direct conversion channel in the multi-channel recording system has led to the use of advanced technology nodes such as 65 nm, 55 nm, 40 nm, and 22 nm as shown in Figs. 2 and 4. But still, the 180 nm node has been widely employed for implementing multi-channel recording systems in GP and MEA systems (Figs. 2 and 4). Also, considering manufacturing costs and circuit prototypes implemented with a small number of channels, designing ICs using the 180 nm node is still an attractive approach. Note that Fig. 4 is generated when information on the technology node and information on the number of active front-end channels are included together in the literature, therefore, some data points are missing in Fig 4 if the literature includes only one of the two.

## IV. TREND OF THE SUPPLY VOLTAGE DRIVING THE ACTIVE FRONT-END CHANNEL

The most straightforward way for reducing the power consumption of each front-end channel is to down the supply voltage and bias current. But the bias current is related to the channel performance, e.g., the input-referred noise and bandwidth. It means that a certain amount of the bias current must flow through the transistors for achieving the low input-referred noise and required bandwidth. Accordingly, minimizing the supply voltage is preferable for reducing the overall power consumption. But the minimum supply voltage must be secured for ensuring the overdrive voltage and output voltage swing.

Fig. 5 presents the trend of the supply voltage used for driving the active front-end channel from the 1970s to the 2020s. From the 1970s to the 1990s, the front-end channels were driven using relatively high supply voltages, which increased the power consumption of each channel. Since the early 2000s, various supply voltages have been used and the overall supply voltage trend has declined. Therefore, the front-end channels have been energy-efficiently designed. In particular, for optimizing the power consumption of the front-end channel composed of analog and digital circuits, multiple supply voltages can be used. Thus, analog circuits that need to ensure overdrive voltages and large voltage swings are driven with relatively high supply voltages while driving digital circuits with lower supply voltages. For further optimizing the power consumption in analog circuits, e.g., multi-stage OTAs, the transistors used in the multi-stage OTA can be driven with different supply voltages according to the required performance. So the transistors are driven with relatively high supply voltages if large voltage swings are required, otherwise the transistors are driven with lower supply voltages.

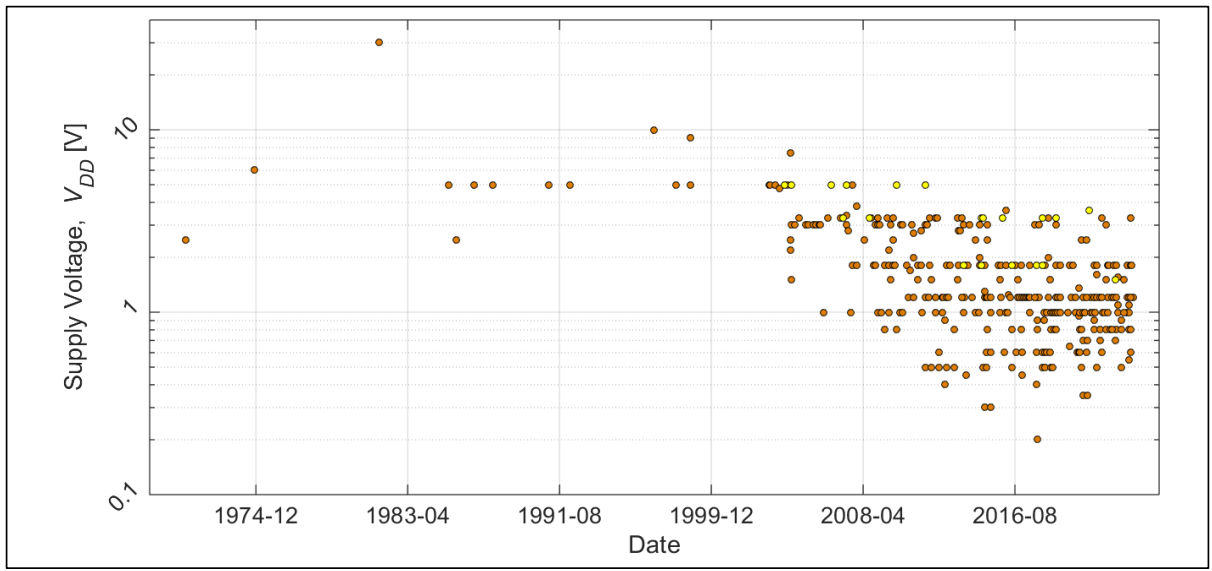

Fig. 5. Trend of the supply voltage $V_{DD}$ driving the active front-end channel from the 1970s to the 2020s.

In the biopotential recording front-end channels including the analog- and digital-intensive topologies, one of the circuit blocks consuming a lot of power is the OTA. So designing the OTA while achieving low power consumption and low input-referred noise is important. For this reason, the inverter-based, also known as the current-reuse, structure is widely employed for the OTA input stage rather than using the PMOS (or NMOS) pair [106], [138]. Thanks to the characteristics of the PMOS and NMOS sharing the same bias current in the inverter-based structure, the current-noise efficiency is improved, resulting in the increased relative transconductance $g_m/I_D$ compared to that of the PMOS (or NMOS) pair. Also, several design techniques that drive the inverter-based stages with lower supply voltages are developed for reducing power consumption.

Many inverter-based OTAs are implemented in two stages to provide a sufficient open-loop gain. Generally, the first input stage is designed to consume most of the bias current since the input stage determines the overall input-referred noise. The second output stage needs to deal with large voltage swings. To design the OTA energy-efficiently while considering these design conditions, the input stage can be driven with lower supply voltages while driving the output stage with relatively high supply voltages. Especially, several design techniques are developed for driving the inverter-based input stages with lower supply voltages while maintaining a certain amount of current consumption for low noise performance [173], [215], and [260]. Under a low supply voltage, the gate terminals of the PMOS and NMOS in the inverter-based stage can be separately DC-biased for ensuring the proper gate-source voltage of each transistor [173], [215], and [260].

As shown in Fig. 5, the supply voltage driving the channel starts to decline since the early 2000s, which is similar to the trend of the technology node shown in Fig. 2. Fig. 6 presents the relationship between technology nodes and supply voltages used for the biopotential recording front-end channels. Approximately, but not absolutely, the front-end channels fabricated using advanced technology nodes tend to be driven with lower supply voltages. Especially, the 180 nm node is widely employed in a variety of supply voltages ranging from a few hundred millivolts to a few volts. The relationship between the number of active front-end channels and supply voltages is investigated as shown in Fig. 7. Most of the multi-channel recording systems with a few hundred channels are implemented using supply voltages greater than one volt, while most of the multi-channel recording systems with a few tens of channels are designed using a variety of supply voltages ranging from a few hundred millivolts to a few volts. Note that the technology node, number of front-end channels, and supply voltage are investigated based on the front-end recording ICs developed from the 1970s to the 2020s but some values in each category are not able to be found in the literature. Therefore, some data points are missing in Figs. 6 and 7. In other words, Fig. 6 is generated when information about the technology node and supply voltage are included together, and Fig. 7 is generated when information about the number of active front-end channels and supply voltage are included together.

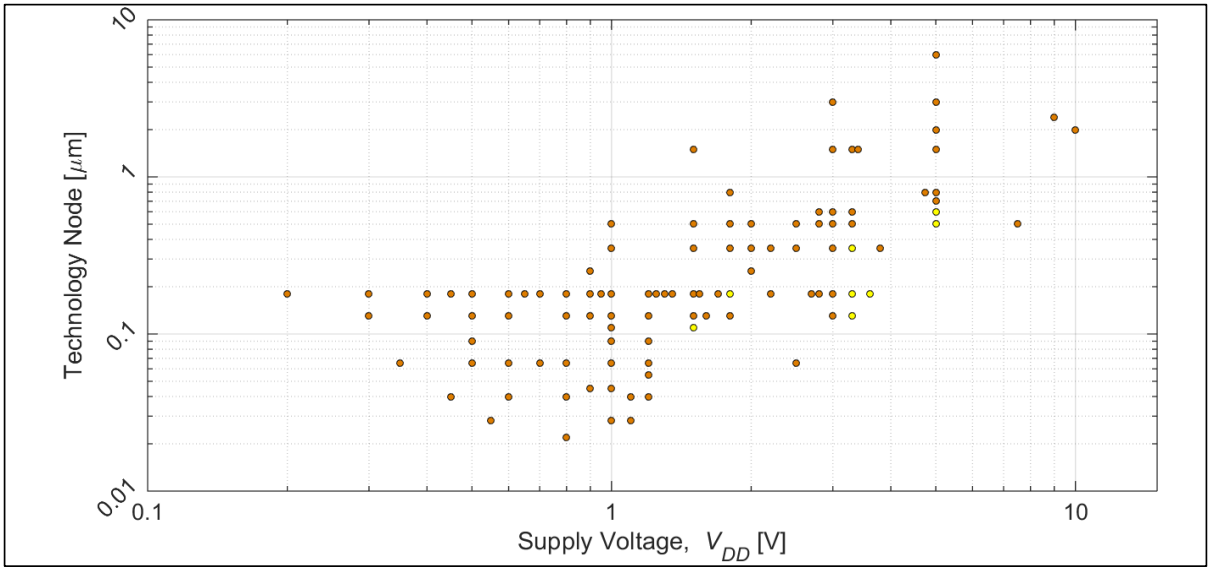

Fig. 6. Distribution of technology nodes according to supply voltages used for the active front-end channels.

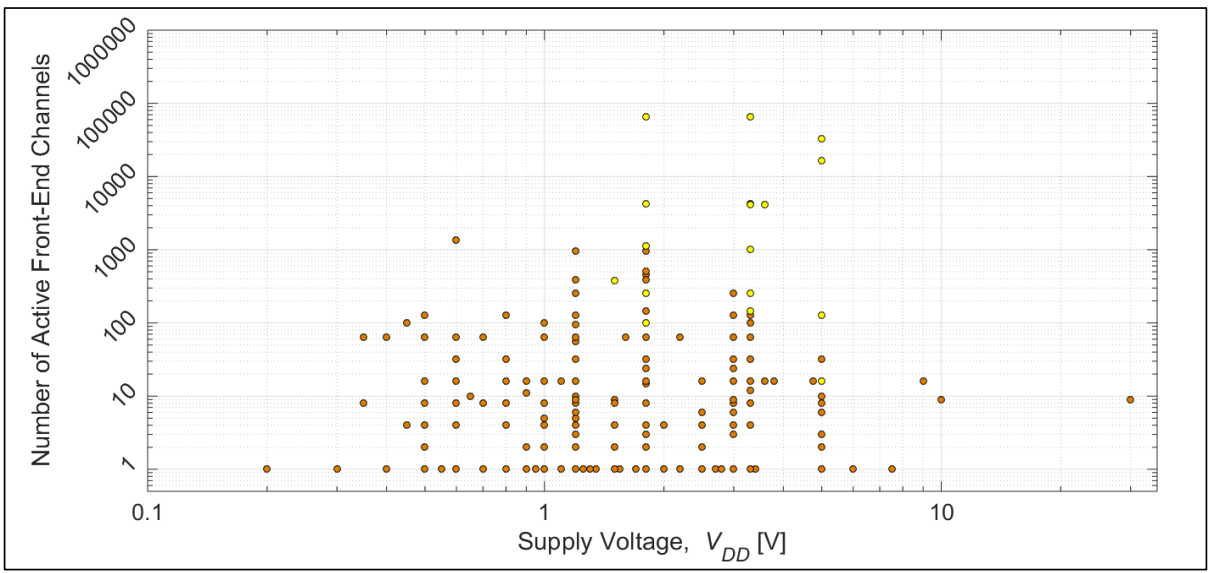

Fig. 7. Distribution of the number of active front-end channels according to supply voltages.

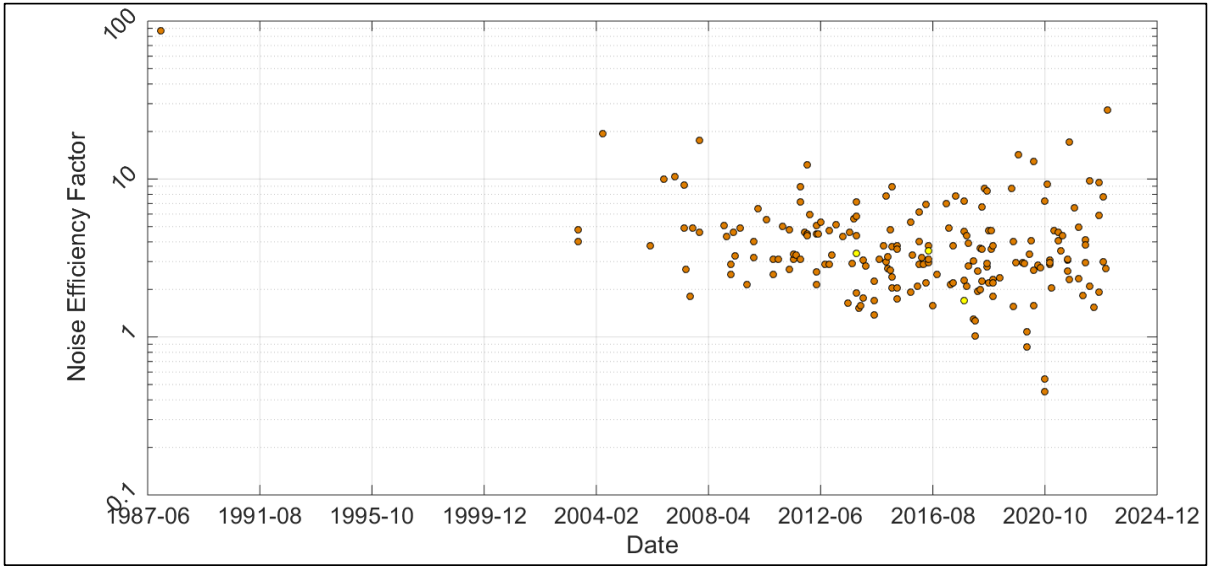

Fig. 8. Trend of the noise efficiency factor (NEF) of the biopotential recording front-end channel.

## V. NOISE EFFICIENCY FACTOR AND POWER EFFICIENCY FACTOR

To evaluate the performance of the biopotential recording front-end channel in terms of the input-referred noise and bias current, the noise efficiency factor (NEF), defined as $V_{n,rms}\sqrt{2I_{tot}/(\pi \cdot U_T \cdot 4kT \cdot BW)}$, has been widely employed [7]. $V_{n,rms}$ is the input-referred noise, $I_{tot}$ is the total bias current consumed by the channel, $U_T$ is the thermal voltage of $kT/q$, $k$ is the Boltzmann constant, $T$ is the absolute temperature, and $BW$ is the channel bandwidth. The NEF is a normalized equation by comparing the input-referred noise of the channel to that of a single bipolar junction transistor (BJT), which can also be expressed as $V_{n,rms}/\sqrt{(\pi \cdot U_T \cdot 4kT \cdot BW)/2I_{tot}}$ when considering only the thermal noise [7]. However, the NEF does not include the supply voltage $V_{DD}$ and only deals with the current-noise efficiency. Therefore, the power-noise efficiency between different recording channels using different $V_{DD}$s can not be fairly compared. For this reason, the power efficiency factor (PEF), defined as $V_{DD} \cdot NEF^2$, is developed [104]. The NEF and PEF have been widely used as the figure of merits comparing the current- and power-noise efficiencies, respectively, between numerous biopotential recording front-end channels. In the general design of the front-end channel composed of the multiple amplification stages such as the LNA and VGA, most of the bias current is consumed by the input stage of the LNA since the input-referred noise voltage is dominantly determined by the devices in the input stage. Other amplification stages following the LNA can be designed with a lower bias current although the noise may increase. Fortunately, the noise of the stages following the LNA is negligible when referred to the LNA input as the noise is divided by the LNA gain. Therefore, the NEF and PEF are approximately dominated by the first stage of the whole biopotential recording channel.

Fig. 8 presents the NEF trend of the biopotential recording front-end channel. Since most of the previous works do not mention the NEF to the early 2000s, the data points during this period are quite missing in Fig. 8. The NEF has been improved through various circuit techniques over time. In the early stage of the development of the biopotential recording front-end channels, the differential pair consisting of the PMOS or NMOS is used as the OTA input stage. As expressed in the NEF definition, $V_{n,rms}$ must be reduced without an increase of $I_{tot}$ for achieving a low NEF. Considering only the thermal noise in a voltage amplifier used for biopotential recording, a high transconductance is required in the OTA input stage to achieve a low $V_{n,rms}$. Therefore, the inverter-based structure is widely employed as an input stage of the voltage amplifier, which boosts the transconductance of the amplifier without a significant increase of $I_{tot}$ [106], [138]. Therefore, the theoretical lower limit of the NEF is reduced when using the inverter-based input stage than that of the PMOS- or NMOS-based input stage. But the design area of the inverter-based input stage increases by two times compared to the PMOS- or NMOS-based input stage. Also, considering that the area of input transistors needs to be widened for reducing the flicker noise, the area increase can be more significant than that of the PMOS- or NMOS-based input stage as the number of front-end channels increases.

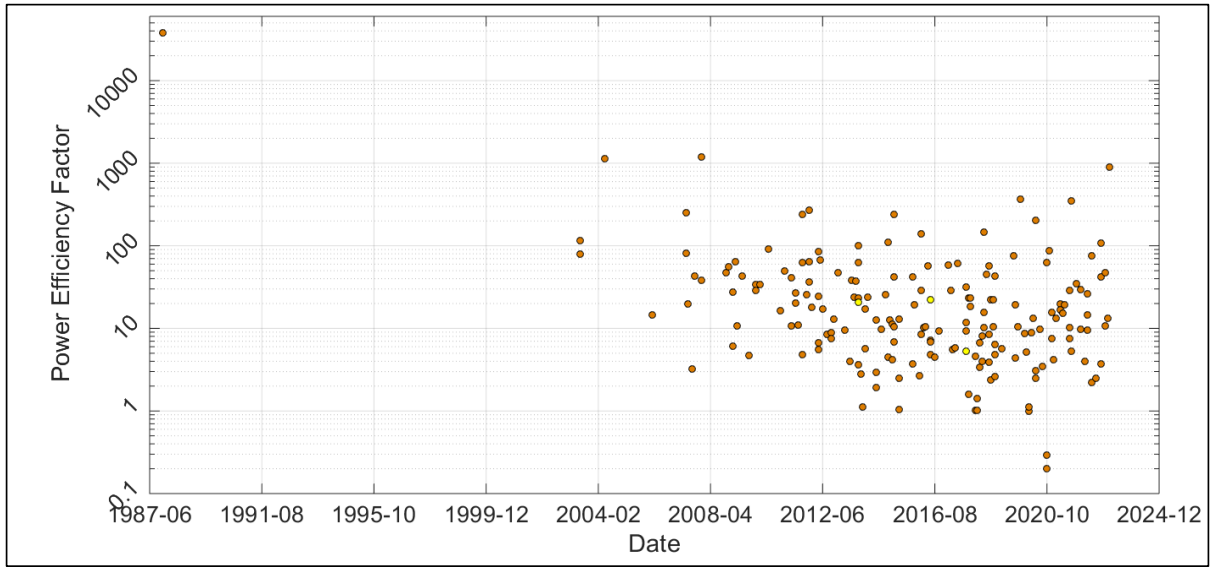

Fig. 9. Trend of the power efficiency factor (PEF) of the biopotential recording front-end channel.

To further reduce the theoretical lower limit of the NEF in the inverter-based input stage, the inverter structure can be vertically stacked in the input stage of the front-end channel [149], [221], and [261]. By employing the inverter-based input stage and the vertically-stacked inverter input stage, the theoretical lower limit of the NEF can be reduced compared to the PMOS- or NMOS-based input stage. However, the DC bias points of the PMOS and NMOS used in the inverter-based input stage need to be carefully set for the proper operation of the front-end channel, especially, when using the low supply voltage and vertically-stacked inverter input stage. As the number of stacked inverters increases, a higher supply voltage can be required to ensure the overdrive voltage of each transistor, which results in increased power consumption despite an improved NEF. The design area of each channel increases when increasing the number of stacked inverters, which significantly increases the overall design area of the multi-channel recording system. Also, the linearity of the channel can be degraded as the number of stacked inverters increases. In addition to the inverter-based designs based on the continuous-time operation, the front-end channel can be designed based on the discrete-time operation to reduce the input-referred noise and consequently, the NEF and PEF are improved [275]. However, the discrete-time-based design can increase the single channel area due to the passive components. Therefore, the proper design techniques must be chosen considering the target design area, number of front-end channels, power consumption, etc.

Fig. 9 presents the PEF trend of the biopotential recording front-end channel. Due to the lack of information on the NEF and PEF in several works, the data points from the 1970s to the early 2000s are quite missing in Fig. 9. If the literature does not mention the PEF while describing the NEF and supply voltage $V_{DD}$, the PEF is calculated as $V_{DD} \cdot \text{NEF}^2$ and included in Fig. 9. Similar to the NEF trend, the PEF has been improved over the entire period thanks to various circuit techniques improving the current-noise efficiency along with lower supply voltages. Especially, some channels having higher NEFs than other channels show better PEFs thanks to the use of lower supply voltages. In other words, the use of lower supply voltages along with circuit techniques, that improve the current-noise efficiency, has led to further improved PEFs. For investigating the NEF, PEF, and supply voltage in each front-end channel, Fig. 10 is generated using the definition of PEF = $V_{DD}·$NEF$^2$. Fig. 10(a) shows PEF versus NEF and Fig. 10(b) presents an enlarged region where the data points are concentrated. The lines in Figs. 10(a) and (b) indicate the supply voltage contour expressed in the log scale. The supply voltage contour is drawn from 0.1 V to 0.9 V with a step of 0.1 V and also drawn from 1 V to 10 V with a step of 1 V.

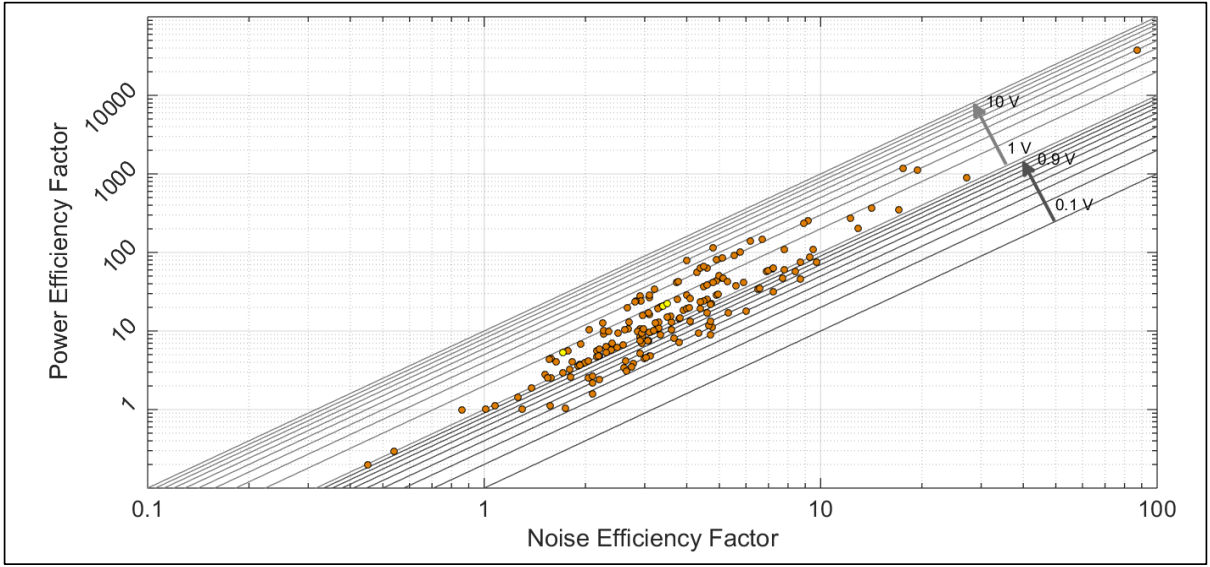

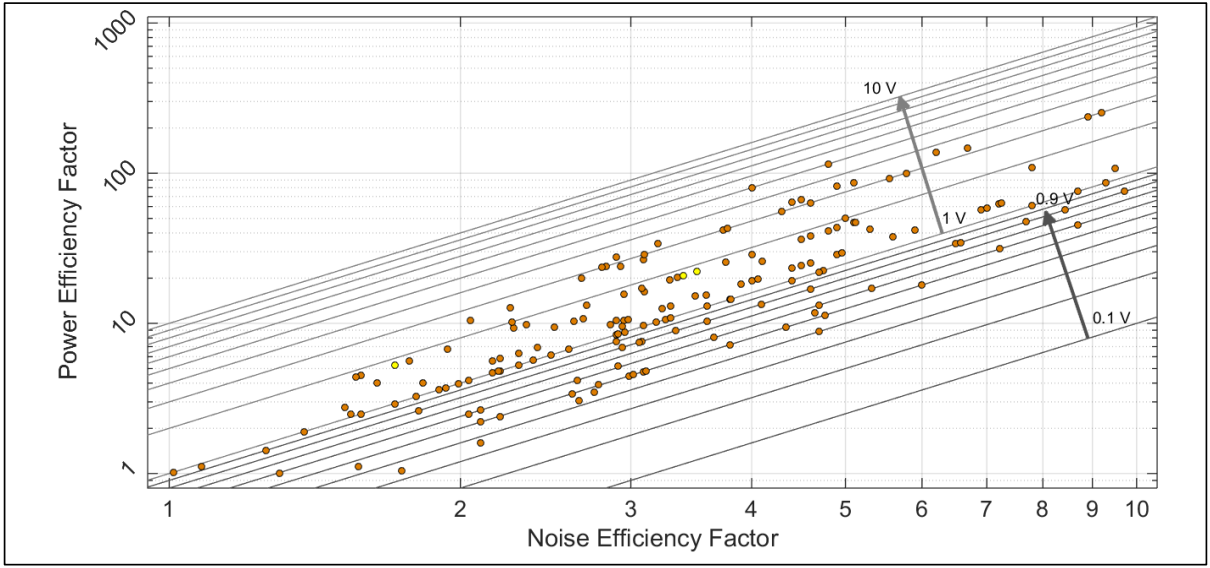

Fig. 10. (a) PEF versus NEF. (b) Enlarged PEF versus NEF (each line means the supply voltage contour).

Figs. 11(a) and (b) present the NEF and PEF distributions according to the technology nodes used in the front-end channels, respectively. Note that some data points below the 65 nm node are missing in Figs. 11(a) and (b) compared to Fig. 2 due to the lack of information in the literature. As discussed in Fig. 2, numerous biopotential recording front-end channels are developed using the 180 nm node, which also can be shown in Figs. 11(a) and (b). Among the various technology nodes, the best performance of both NEF and PEF is achieved using the 180 nm node. However, the NEF and PEF achieved using the sub-65 nm nodes do not ensure the best performance due to the high input-referred noise. When looking at the NEF and PEF distributions achieved using the 180 nm node, the NEFs range from less than one to several tens while the PEFs range from less than one to several hundreds. When converting the NEF to the PEF, the NEF is squared and multiplied by $V_{DD}$. Thus, the PEF increases substantially if the NEF is greater than 1 and $V_{DD}$ is equal to or greater than 1 V. Fortunately, the increase in the PEF can be mitigated even with a high $V_{DD}$ when the NEF is achieved as less than 1.

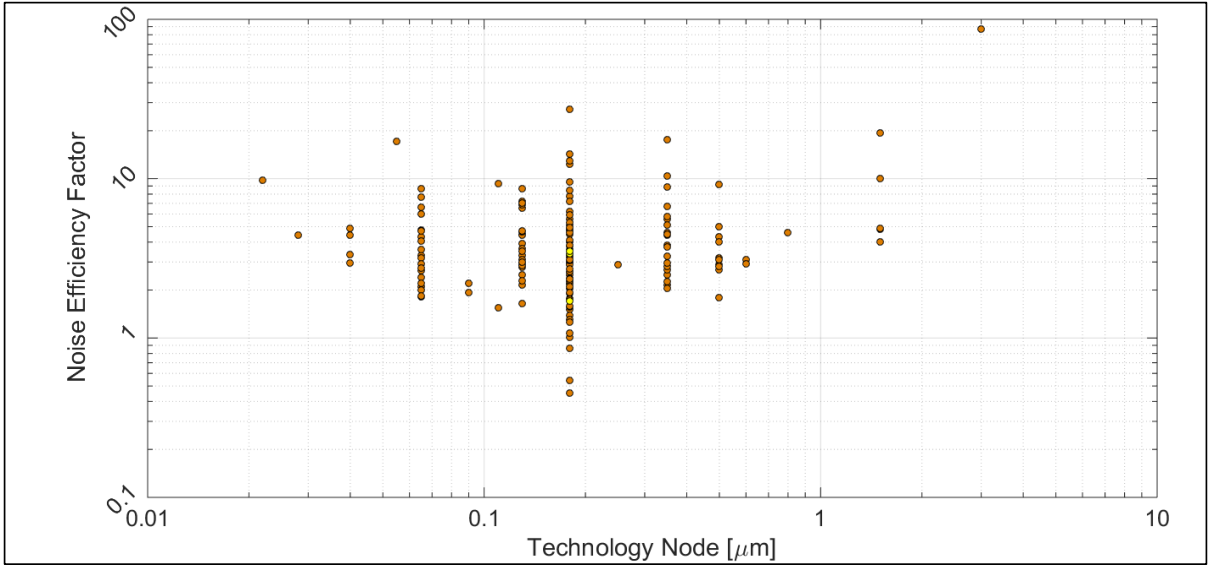

(a)

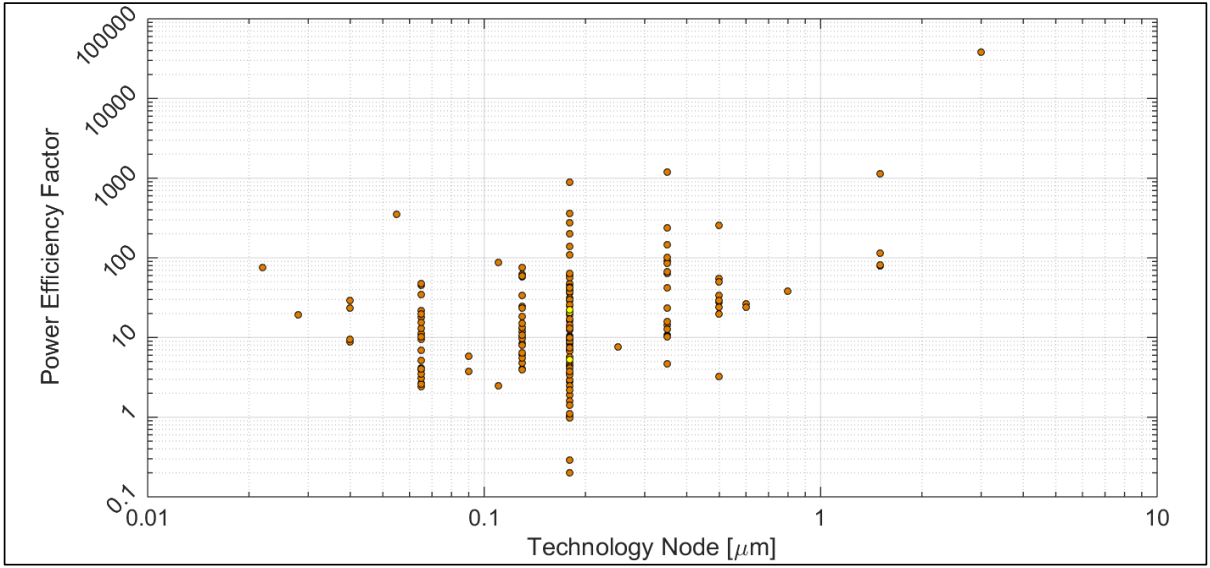

(b)

Fig. 11. (a) NEF distribution and (b) PEF distribution according to technology nodes.

Figs. 12(a) and (b) show the NEF and PEF distributions according to the number of front-end channels, respectively. Note that since most of the MEA systems do not mention the NEF and PEF, data points are considerably missing in Figs. 12(a) and (b). The NEF and PEF can be improved by using circuit techniques such as vertically-stacked inverters, discrete-time operation, etc. However, these techniques require additional transistors and passive components, thereby increasing each channel area and eventually limiting the number of front-end channels for the high-density multi-channel recording systems. For this reason, the achievable NEF and PEF may be compromised depending on the design area and number of channels.

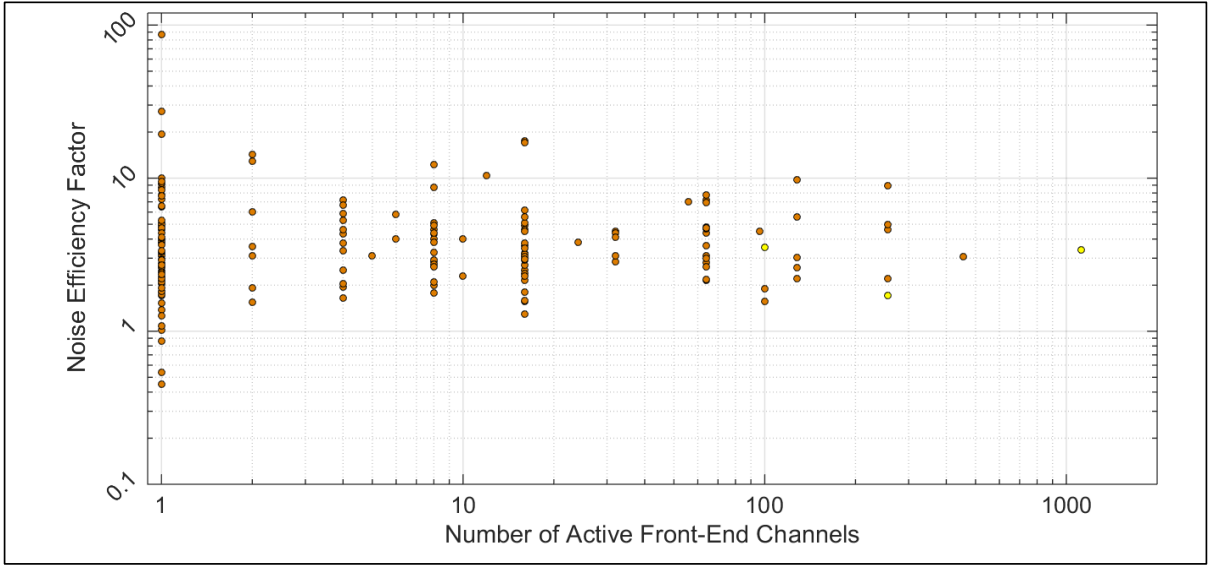

(a)

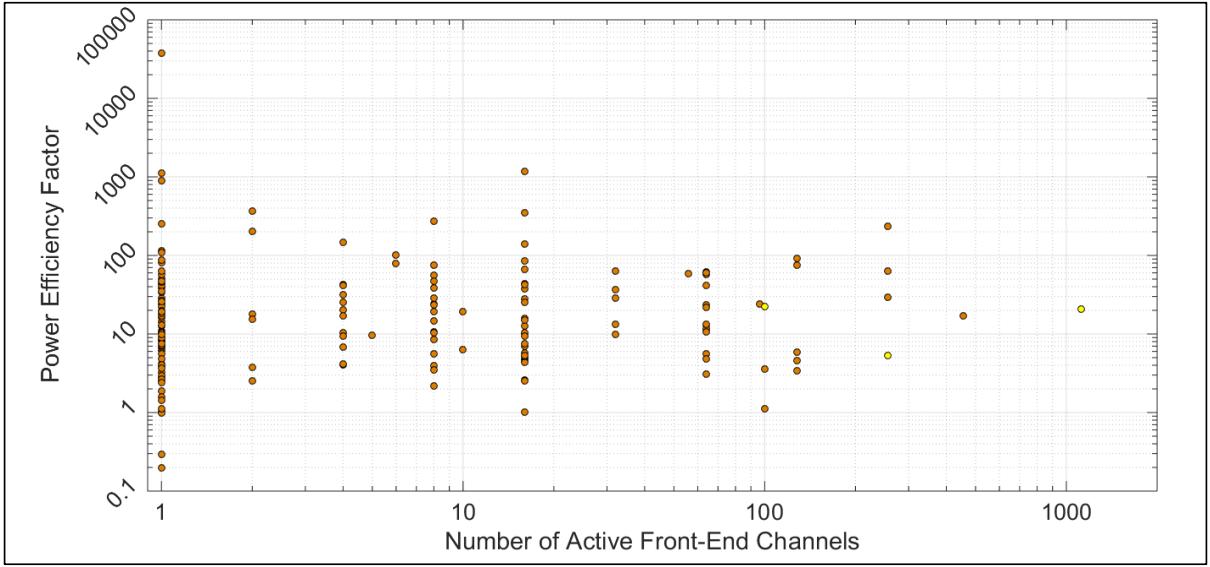

(b)

Fig. 12. (a) NEF distribution and (b) PEF distribution according to the number of active front-end channels.

VI. CONCLUSION

Over the past decades, biopotential recording front-end channels have evolved along with the advances in semiconductor technology. The front-end channels are designed to have appropriate functions depending on whether they are used invasively or non-invasively. However, many challenges still need to be overcome to design a front-end channel suitable for each purpose of use. Especially, when considering invasive recording applications, the number of channels must increase even more dramatically for understanding the broad region of the brain in primates and eventually in humans. While increasing the number of channels to receive a large amount of information, safety issues for subjects must be addressed. An invasive recording is a good means to obtain high-quality signals from neurons although it entails difficulties such as surgery, immune response, long-term monitoring, etc. Therefore, a non-invasive recording system can be a way to avoid the safety issues of invasive recording. However, the poor signal quality of a non-invasive recording limits signal processing compared to an invasive recording. Therefore, even if a signal is recorded non-invasively, techniques for reconstructing the signal to a level comparable to that observed through invasive recording must be developed, which will enable efficient biopotential recording while appropriately supplementing the disadvantages of the two methods.

Through the investigation of the front-end channels developed from the 1970s to the 2020s, it is found that various circuit techniques have been developed to improve the performance in channel density, current-noise efficiency, power-noise efficiency, signal quality, etc. But no front-end channel achieves the best performance in all aspects. Trade-offs are always entailed when employing a circuit technique for improving a specific performance. For this reason, the appropriate circuit techniques need to be selected with some compromise in the performance according to the purpose of use. Since the 1970s, the technology node has evolved by gradually shrinking, and front-end channels have been developed using diverse technology nodes depending on the purpose of use while considering the performance and fabrication costs. Numerous front-end channels have been developed using standard CMOS technology. Also, HV-CMOS and BCD technologies have been widely used for designing the recording channels in conjunction with the stimulation circuits in neuromodulation applications. Considering the immense advancement of the biopotential recording front-end channels during the last decades, we forecast an optimistic future that enables neural network analysis even more extensively than the prior studies.